# Apparent measurement errors in "Development of biomechanical response corridors of the thorax to blunt ballistic impacts"

Michael Courtney[a], Amy Courtney[b]
[a]Ballistics Testing Group, P.O. Box 24, West Point, NY 10996
email: Michael_Courtney@alum.mit.edu
[b]Department of Physics, United States Military Academy, West Point, NY 10996

**Abstract:** "Development of biomechanical response corridors of the thorax to blunt ballistic impacts" (Bir, C., Viano, D., King, A., 2004, Journal of Biomechanics 37, 73-79.) contains apparent measurement errors. Areas under several force vs. time (Fig. 2) and force vs. deflection curves (Fig.4) differ significantly from the momentum and kinetic energy changes, respectively.

*Keywords:* Thoracic response; Impact biomechanics; Ballistic; Blunt trauma

Studying detailed response curves is an important contribution to understanding blunt ballistic impacts (Bir, C. Viano, D., King, A., 2004, Development of biomechanical response corridors of the thorax to blunt ballistic impacts, Journal of Biomechanics 37, 73-79.) Newton's second law demands the integral of force vs. time equal momentum change. The work-energy theorem demands the integral of force vs. distance equal change in kinetic energy. For a projectile coming to rest, integrals of force vs. time and force vs. distance should be equal in magnitude to the initial momentum and the initial kinetic energy of the projectile, respectively.

Figures 2 and 4 **(Bir et al., 2004)** show measured curves for force vs. time and force vs. deflection, respectively. The shape of these response curves depends on interactions between projectile and target. However, if accurately measured, the area under each response curve should depend only on the momentum and energy of the projectile.

Initial momenta of projectiles whose response curves are shown in Fig. 2 (a) and (b) are 2.8 kg m/s and 5.6 kg m/s, respectively. Inspection shows that areas under several curves differ significantly from expected values. Variation is roughly a factor of two in the area under the different curves for a given sub-figure. This variation indicates measurement error, because curves represent different trials with the same momentum change.

Initial energies for response curves in Fig. 4 (a) and (b) are 28 J and 112 J, respectively. Areas under several curves differ significantly from expected values. Variation in the area under different curves of each sub-figure exceeds a factor of two. This variation indicates measurement error, because different trials had the same kinetic energy change.

Possible sources of discrepancy include variations in projectile velocity, miscalibration of accelerometers, and projectile rotation confounding acceleration measurements. If the acceleration measurement system was miscalibrated and the response is linear, the data can be reinterpreted by normalizing response curves so that the areas have the values expected from momentum and kinetic energy considerations. This would lead to a narrowing of the reported response corridors.